\documentclass[12pt]{article}
\usepackage{amsfonts}
\usepackage{amsmath}
\usepackage{amssymb}

\input{tcilatex}
\begin{document}

\begin{center}
\smallskip \ 

\textbf{GENERALIZED ELKO THEORY}

\smallskip \ 

\smallskip \ 

J. A. Nieto \footnote{%
nieto@uas.edu.mx, janieto1@asu.edu}

\smallskip \ 

\smallskip \ 

\smallskip \ 

\textit{Facultad de Ciencias F\'{\i}sico-Matem\'{a}ticas de la Universidad
Aut\'{o}noma} \textit{de Sinaloa, 80010, Culiac\'{a}n Sinaloa, M\'{e}xico.}

\smallskip \ 

\smallskip \ 

\textbf{Abstract}
\end{center}

By using a totally antisymmetric spinor field we generalize Elko theory. We
compare our proposed theory with traditional totally antisymmetric spinor
field theory based on the Dirac equation. As an application of our formalism
we comment about the possibility to link our generalized Elko theory with
matroids, qubits and surreal numbers.

\bigskip \ 

\bigskip \ 

\bigskip \ 

\bigskip \ 

\bigskip \ 

\bigskip \ 

\bigskip \ 

\bigskip \ 

\bigskip \ 


Pacs numbers: 04.20.Gz, 04.60.-Ds, 11.30.Ly

June, 2019

\newpage

\noindent It is known that the Dirac equation [1] can be considered as the
`square root' of the Klein-Gordon equation. Surprisingly, in 2005 Ahluwalia
and Grumiller [2]-[3] (see also [4]-[7]) proved that such a `square root' it
is not unique. In fact, assuming a more general helicity eigenvalues they
proved that an alternative and different field equation emerges which can
also be reduced to the Klein-Gordon equation. They called their formalism
Elko theory [2]-[3] (see also Refs [8]-[22]). It turns out that the physical
states of the Elko equation can also be identified with $\frac{1}{2}$-spin
fermions, but with the property that such a fermions are non-charged. One of
the interesting aspects of the Elko fermions is that they provide one of the
most interesting candidate for dark matter [23].

In this note we propose a generalized Elko equation which also implies the
Klein-Gordon equation. The physical states of our proposed equation are $l$%
-rank totally antisymmetric tensor spinor fields of the form $\psi
_{aA_{1}...A_{l}}(x^{\mu })$, which are different from the common totally
antisymmetric spinor fields $\psi _{a\mu _{1}...\mu _{l}}(x^{\mu })$
satisfying the Dirac equation [24]-[26]. An interesting feature of our
formalism arises when one requires that the field $\psi _{aA_{1}...A_{l}}$
satisfies the Grassmann-Pl\"{u}cker relations. In this case $\psi
_{aA_{1}...A_{l}}$ becomes a decomposable tensor and we mention that this
result establishes a connection no only with matroid theory but also with
qubits and surreal number theory.

Let us start considering the fundamental equation in which the Elko theory
[2]-[3] is based, namely

\begin{equation}
\left[ \gamma _{ab}^{\mu }\hat{p}_{\mu }\delta _{A}^{B}+im_{0}\delta
_{ab}\varepsilon _{A}^{~~B}\right] \psi _{B}^{b}=0.  \label{1}
\end{equation}%
Here, the indices $\mu ,\nu $ etc. run from $0$ to $3$, the indices $A,B$
run from $1$ to $2$ and the spinorial indices $a,b$ run from $1$ to $4$.
Further, the $\gamma $-matrices satisfy the Clifford algebra

\begin{equation}
\gamma ^{\mu }\gamma ^{\nu }+\gamma ^{\nu }\gamma ^{\mu }=2\varepsilon \eta
^{\mu \nu },  \label{2}
\end{equation}%
with $\eta ^{\mu \nu }=diag(-1,1,1,1)$ and $\varepsilon =\pm 1$. Moreover, $%
\hat{p}_{\mu }=-i\partial _{\mu }$ and $\varepsilon ^{AB}$ is the totally
antisymmetric $\varepsilon $-symbol, with $\varepsilon ^{12}=1=-\varepsilon
^{21}$.

In contrast to the usual Dirac equation [1]

\begin{equation}
\left[ \gamma _{ab}^{\mu }\hat{p}_{\mu }+m_{0}\delta _{ab}\right] \psi
^{b}=0,  \label{3}
\end{equation}%
equation (1) requires $8$-component complex spinor field $\psi _{A}^{a}$
rather than $4$-component $\psi ^{a}$ which is the case in equation (3).
Furthermore, the quantities $\delta _{A}^{B}$ and $\varepsilon _{A}^{~~B}$
in (1) establish that $\psi _{A}^{a}$ is not eigenspinor of the $\gamma
^{\mu }\hat{p}_{\mu }$ operator as $\psi ^{a}$ in the equation (3). In spite
of these key differences, one can prove that both field equations (1) and
(3) imply the Klein-Gordon equations

\begin{equation}
\left[ \hat{p}^{\mu }\hat{p}_{\mu }+m_{0}^{2}\right] \psi _{A}^{a}=0
\label{4}
\end{equation}%
and%
\begin{equation}
\left[ \hat{p}^{\mu }\hat{p}_{\mu }+m_{0}^{2}\right] \psi ^{a}=0,  \label{5}
\end{equation}%
respectively. Therefore, both (1) and (3) can be understood as the `square
root' of the Klein-Gordon equation. This proves that such a `square root' is
not unique.

There are at least two important remarks that emerge from (1). One related
to the `background' space associated with the indices $\mu ,\nu $...etc. and
the other associated with an `internal' space associated with the indices $%
A,B$...etc. In the first case, higher dimensional background space, with the
indices $\hat{\mu},\hat{\nu}$ ...etc. replacing the indices $\mu ,\nu $ etc.
and running from $0$ to $d-1$, generalizes (1) in the form

\begin{equation}
\left[ \gamma _{ab}^{\hat{\mu}}\hat{p}_{\hat{\mu}}\delta
_{A}^{B}+im_{0}\delta _{ab}\varepsilon _{A}^{~~B}\right] \psi _{B}^{b}=0,
\label{6}
\end{equation}%
which implies the $d$-dimensional Klein-Gordon equation

\begin{equation}
\left[ \hat{p}^{\hat{\mu}}\hat{p}_{\hat{\mu}}+m_{0}^{2}\right] \psi
_{B}^{a}=0.  \label{7}
\end{equation}%
Of course, one must also generalize $\gamma $-matrices $\gamma ^{\mu
}\rightarrow \gamma ^{\hat{\mu}}$ in the Clifford algebra (2) and the spinor
field $\psi _{B}^{a}$ must increase its components as a power of $2^{\frac{d%
}{2}}$.

The second case, it arises from the observation that the combination $%
\varepsilon ^{AB}\psi _{B}^{a}$ can be interpreted as the dual $^{\ast }\psi
^{aA}$ of $\psi _{B}^{a}$. In fact, one may define

\begin{equation}
^{\ast }\psi ^{aA}\equiv \varepsilon ^{AB}\psi _{B}^{a}.  \label{8}
\end{equation}%
This means that (1) can also be written as 
\begin{equation}
\gamma _{ab}^{\mu }\hat{p}_{\mu }\psi _{A}^{b}+im_{0}\text{ }\delta
_{ab}~^{\ast }\psi _{A}^{b}=0.  \label{9}
\end{equation}%
So, one has discovered that (1) or (9) combines $\psi _{A}^{a}$ and $^{\ast
}\psi _{A}^{a}$ in a particular way. An interesting observation, that we
have not seen in the literature, it is that since $^{\ast \ast }\psi
_{A}^{a}=-\psi _{A}^{a}$ one finds that the equation (9) implies%
\begin{equation}
\gamma _{ab}^{\mu }\hat{p}_{\mu }~^{\ast }\psi _{A}^{b}-im_{0}\delta _{ab}%
\text{ }\psi _{A}^{b}=0.  \label{10}
\end{equation}

Looking (8) and knowing that duality is central concept in totally
antisymmetric tensors one is motivated to introduce the spinorial field $%
\psi _{B_{1}...B_{l}}^{a}$ and its dual

\begin{equation}
^{\ast }\psi ^{aA_{1}...A_{l}}=\frac{1}{l!}\varepsilon
^{A_{1}...A_{l}B_{1}...B_{l}}\psi _{B_{1}...B_{l}}^{a}.  \label{11}
\end{equation}%
Here, $\varepsilon ^{A_{1}...A_{l}B_{1}...B_{l}}$ is the totally
antisymmetric Levi-Civita symbol ($\varepsilon $-symbol) and the indices $%
A,B $ run, in this case, from $1$ to $2l$, with $l$ a natural number. Now
one may ask; what is it the Elko equation for the field $\psi
_{B_{1}...B_{l}}^{a}$?

Our proposed generalized Elko equation is

\begin{equation}
\left[ \frac{1}{l!}\gamma _{ab}^{\mu }\hat{p}_{\mu }\delta
_{A_{1}...A_{l}}^{B_{1}...B_{l}}+\frac{i}{l!}m_{0}\delta _{ab}\varepsilon
_{A_{1}...A_{l}}^{~~~~~~B_{1}...B_{l}}\right] \psi _{B_{1}...B_{l}}^{a}=0,
\label{12}
\end{equation}%
where $\delta _{A_{1}...A_{l}}^{B_{1}...B_{l}}$ is the generalized Kronecker
delta which is related to $\varepsilon
_{A_{1}...A_{l}}^{~~~~~~B_{1}...B_{l}} $ by

\begin{equation}
\delta _{A_{1}...A_{l}}^{B_{1}...B_{l}}=\frac{1}{l!}\varepsilon
_{C_{1}...C_{l}}^{~~~~~~B_{1}...B_{l}}\varepsilon
_{~~~~~~A_{1}...A_{l}}^{C_{1}...C_{l}}.  \label{13}
\end{equation}%
Considering the computation

\begin{equation}
\begin{array}{c}
\left[ \frac{1}{l!}\gamma _{a}^{\mu ~c}\hat{p}_{\mu }\delta
_{A_{1}...A_{l}}^{C_{1}...C_{l}}-\frac{i}{l!}m_{0}\delta _{a}^{c}\varepsilon
_{A_{1}...A_{l}}^{~~~~~~C_{1}...C_{l}}\right] \times \\ 
\\ 
\times \left[ \frac{1}{l!}\gamma _{cb}^{\nu }\hat{p}_{\nu }\delta
_{C_{1}...C_{l}}^{B_{1}...B_{l}}+\frac{i}{l!}m_{0}\delta _{cb}\varepsilon
_{C_{1}...C_{l}}^{~~~~~~B_{1}...B_{l}}\right] \psi _{B_{1}...B_{l}}^{a}=0,%
\end{array}
\label{14}
\end{equation}%
and using (13) one obtains that (14) implies the equation

\begin{equation}
\left[ \frac{1}{l!}\gamma _{a}^{\mu ~c}\gamma _{cb}^{\nu }\hat{p}_{\mu }\hat{%
p}_{\nu }\delta _{A_{1}...A_{l}}^{B_{1}...B}+\frac{\varepsilon }{l!}%
m_{0}^{2}\delta _{A_{1}...A_{l}}^{B_{1}...B_{l}}\right] \psi
_{B_{1}...B_{l}}^{a}=0  \label{15}
\end{equation}%
which in virtue of (2) leads to the Klein-Gordon equation

\begin{equation}
\left[ \hat{p}^{\mu }\hat{p}_{\mu }+m_{0}^{2}\right] \psi
_{A_{1}...A_{l}}^{a}=0.  \label{16}
\end{equation}%
The factor $\varepsilon $ in (15) is due to the permutation order in the
definition (13), that is, according to (13) $\varepsilon =+1$ if $l$ is even
and $\varepsilon =-1$ if $l$ is odd.

Just as $\psi _{A}^{a}$ is not eigenspinor of the $\gamma ^{\mu }\hat{p}%
_{\mu }$ operator as $\psi ^{a}$ in the equation (3) our proposed equation
(12) also implies that the field $\psi _{A_{1}...A_{l}}^{a}$ is not
eigenspinor of the $\gamma ^{\mu }\hat{p}_{\mu }$ operator in contrast to
the conventional assumption of the totally antisymmetric spinor field in
which the field $\psi _{\mu _{1}...\mu _{l}}^{a}$ is required to satisfy the
Dirac type equation%
\begin{equation}
\left[ \gamma _{ab}^{\mu }\hat{p}_{\mu }+m_{0}\delta _{ab}\right] \psi
_{\alpha _{1}...\alpha _{l}}^{b}=0  \label{17}
\end{equation}%
and the two constrains;

\begin{equation}
\hat{p}^{\alpha _{1}}\psi _{\alpha _{1}...\alpha _{l}}^{b}=0  \label{18}
\end{equation}%
and

\begin{equation}
\gamma ^{\alpha _{1}}\psi _{\alpha _{1}...\alpha _{l}}^{b}=0.  \label{19}
\end{equation}

In order to reduce the number of degrees of freedom of $\psi _{A}^{a}$ from $%
8$-complex components to $4$-complex components in Elko theory a kind of
Majorana condition for the physical states $\psi _{A}^{a}$ is imposed,
namely [2]-[3]%
\begin{equation}
C_{~~b}^{a}\psi _{A}^{b}=e^{i\theta }\psi _{A}^{a},  \label{20}
\end{equation}%
where $C$ denotes a charge conjugation operator and $e^{i\theta }$ is a
phase factor. If one chooses $C_{~~b}^{a}\psi _{A}^{b}=\psi _{A}^{a}$ one
obtains the the self-conjugate Elko spinor or Majorana spinor, while if one
requires the condition $C_{~~b}^{a}\psi _{A}^{b}=-\psi _{A}^{a}$ one gets
the anti-self-conjugate Elko spinor, which is different than the
conventional Majorana choice. The solution of this conditions, in the Weyl
representation, can be written as

\begin{equation}
\psi _{1}^{a}\longrightarrow \lambda =\left( 
\begin{array}{c}
-\sigma ^{2}\psi _{L}^{\ast } \\ 
\psi _{L}%
\end{array}%
\right)  \label{21}
\end{equation}%
and

\begin{equation}
\psi _{2}^{a}\longrightarrow \rho =\left( 
\begin{array}{c}
\psi _{R} \\ 
\sigma ^{2}\psi _{R}^{\ast }%
\end{array}%
\right) .  \label{22}
\end{equation}%
In our generalized Elko equation this analysis is more complicated since $%
\psi _{A_{1}...A_{l}}^{a}$ contains very much degrees of freedom. In fact,
one obtains that $\psi _{A_{1}...A_{l}}^{a}$ contains $\frac{4((2l)!)}{%
(l!)(l!)}$-complex components. Nevertheless, one may write the analogue of
(18) for $\psi _{A_{1}...A_{l}}^{a}$, namely

\begin{equation}
C_{~~b}^{a}\psi _{A_{1}...A_{l}}^{b}=e^{i\theta }\psi _{A_{1}...A_{l}}^{a}.
\label{23}
\end{equation}%
An additional possible constraints arises from the observation that the
constraints (18) and (19) can not be imposed in the field $\psi
_{A_{1}...A_{l}}^{a}$ because the indices $A_{1},...,A_{l}$ are not
spacetime indices as $\alpha _{1}...\alpha _{l}$ in $\psi _{\alpha
_{1}...\alpha _{l}}^{b}$. So, recalling that for a totally antisymmetric
object one can consider the Grassmann-Pl\"{u}cker relations one is motivated
to assume such a relations for each value of the index $a$ in $\psi
_{A_{1}...A_{l}}^{a}$. This means that one may assume that $\psi
_{A_{1}...A_{l}}^{a}$ satisfies the relations:

\begin{equation}
\psi _{A_{1}...[A_{l}}^{a}\psi _{B_{1}...B_{l}]}^{a}=0,  \label{24}
\end{equation}%
for each fixed value of the index $a$. Here, the bracket $%
[A_{l}B_{1}...B_{l}]$ means totally antisymmetric. It is well known that a
totally antisymmetric tensor satisfies the Grassmann-Pl\"{u}cker relations
if and only if it is decomposable. Thus, one finds that due to (24) the
field $\psi _{A_{1}...A_{l}}^{a}$ is decomposable and therefore can be
written as

\begin{equation}
\psi _{A_{1}...A_{l}}^{a}=\Omega _{a_{1}...a_{l}}^{a}\varepsilon
_{i_{1}...i_{l}}\psi _{A_{1}}^{a_{1}i_{1}}...\psi _{A_{l}}^{a_{l}i_{l}},
\label{25}
\end{equation}%
where the only non-vanishing terms of $\Omega _{a_{1}...a_{l}}^{a}$ are $%
\Omega _{1...1}^{1}=\Omega _{2...2}^{2}=\Omega _{3...3}^{3}=\Omega
_{4...4}^{4}=1$. One recognizes in (25) a kind of generalization of the Pl%
\"{u}cker coordinates for $\psi _{A}^{ai}$. Hence, (25) determines one to
one correspondence between the fields $\psi _{A_{1}...A_{l}}^{a}$ and $\psi
_{A}^{ai}$. It turns out that the Pl\"{u}cker coordinates are a key concept
in one of the possible definitions of realizable oriented matroids [27] (see
also Refs. [28]-[36] and references therein). In fact, in this mathematical
scenario one can identify $\psi _{A_{1}...A_{l}}^{a}$ with the complex
version of the chirotopes which are called phirotopes (see Ref. [32] and
references therein) This connection establishes a bridge between our
generalized Elko theory and the mathematical structure of oriented matroids
which can provide with a rich number of possible interesting new physical
routes. In particular, the duality mathematical notion is a central concept
in oriented matroid theory, in the sense that every oriented matroid has a
dual, and therefore the duality expression (11) may also emerges as a key
concept in our proposed generalized Elko theory. It turns out that recently
we have found [36] that the Pl\"{u}cker coordinates is also a key concept in
qubits, twistors and surreal numbers. This means that the expression (25)
leads also to consider our generalized Elko theory in the context of qubit
theory and surreal numbers. Let us briefly explain this observation. In
qubit theory a general a complex state $\mid \psi >\in C^{2^{N}}$ is
expressed as (see Ref. [37] and references therein)

\begin{equation}
\mid \psi >=\dsum \limits_{\hat{A}_{1},\hat{A}_{2},...,\hat{A}_{N}=0}^{1}Q_{%
\hat{A}_{1}\hat{A}_{2}...\hat{A}_{N}}\mid \hat{A}_{1}\hat{A}_{2}...\hat{A}%
_{N}>,  \label{26}
\end{equation}%
where the states $\mid \hat{A}_{1}\hat{A}_{2}...\hat{A}_{N}>=\mid \hat{A}%
_{1}>\otimes \mid \hat{A}_{2}>...\otimes \mid \hat{A}_{N}>$ correspond to a
standard basis of the $N$-qubit. It turns out that, in a particular subclass
of $N$-qubit entanglement, the Hilbert space can be broken into the form $%
C^{2^{N}}=C^{L}\otimes C^{l}$, with $L=2^{N-n}$ and $l=2^{n}$. Such a
partition allows a geometric interpretation in terms of the complex
Grassmannian variety $Gr(L,l)$ of $l$-planes in $C^{L}$ \textit{via} the Pl%
\"{u}cker embedding. The idea is to associate the first $N-n$ and the last $%
n $ indices of $Q_{\hat{A}_{1}\hat{A}_{2}...\hat{A}_{N}}$ with a $L\times l$
matrix $\mathbf{\omega }_{A}^{i}$. This can be interpreted as the
coordinates of the Grassmannian $Gr(L,l)$ of $l$-planes in $C^{L}$. Using
the matrix $\mathbf{\omega }_{A}^{i}$ one can define the Pl\"{u}cker
coordinates

\begin{equation}
\mathcal{Q}_{A_{1}...A_{l}}=\varepsilon _{i_{1}...i_{l}}\mathbf{\omega }%
_{A_{1}}^{i_{1}}...\mathbf{\omega }_{A_{l}}^{i_{l}},  \label{27}
\end{equation}%
which one can associate with (25) by making the identification $\mathbf{%
\omega }_{A}^{i}\rightarrow \psi _{A}^{ai}$. Just as in qubit theory the
transformation $\omega \rightarrow S\omega $, with $S\in GL(l,C)$, the Pl%
\"{u}cker coordinates transform as $\mathcal{Q}\rightarrow Det(S)\mathcal{Q}$
(see Ref. [37] for details) one discovers that under $\psi
_{A}^{ai}\rightarrow S_{j}^{i}\psi _{A}^{aj}$ the field $\psi
_{A_{1}...A_{l}}^{a}$ transform as $\psi _{A_{1}...A_{l}}^{a}\rightarrow
Det(S)\psi _{A_{1}...A_{l}}^{a}$.

Surprisingly, the rebits (the real part of the complex qubits) can be
connected with the fascinating subject of surreal numbers and therefore one
can also associate the Pl\"{u}cker coordinates of $\psi _{A_{1}...A_{l}}^{a}$
with the surreal numbers. There are two equivalent ways for defining surreal
numbers; The Conway [38]-[39] and Gonshor [40] formalisms. Here, although we
shall focus in the Gonshor approach, let us just mention that using Conway
algorithm one finds that at the `$j$-day' one obtains $2^{j+1}-1$ numbers
all of which are of form%
\begin{equation}
x=\frac{p}{2^{q}},  \label{28}
\end{equation}%
where $p$ and $q$ are integers numbers. Of course, the numbers (28) are
dyadic rationals which are dense in the real $R$.

The Gonshor definition of surreal numbers is

\smallskip \ 

\textbf{Definition 1}. A surreal number is a function $f$ from initial
segment of the ordinals into the set $\{+,-\}$.

\smallskip \ 

For instance, if $f$ is the function so that $f(1)=+$, $f(2)=-$, $f(3)=-$, $%
f(4)=+$ then $f$ is the surreal number $(++-+)$. In the Gonshor approach one
obtains the sequence: $1$-day

\begin{equation}
-1=(-),\text{ \  \  \  \  \  \  \  \  \  \  \ }(+)=+1,  \label{29}
\end{equation}%
in the $2$-day

\begin{equation}
-2=(--),\text{ \ }-\frac{1}{2}=(-+),\text{\  \  \ }(+-)=+\frac{1}{2},\text{\  \
\  \  \  \ }(++)=+2,  \label{30}
\end{equation}%
and $3$-day

\begin{equation}
\begin{array}{c}
-3=(---),\text{ \ }-\frac{3}{2}=(--+),\text{\  \ }-\frac{3}{4}=(-+-),\text{\ }%
-\frac{1}{4}=(-++) \\ 
\\ 
(+--)=+\frac{1}{4},\text{ \ }(+-+)=+\frac{3}{4},\text{ \  \ }(++-)=+\frac{3}{2%
},\text{\  \  \  \  \  \ }(+++)=+3,%
\end{array}
\label{31}
\end{equation}%
respectively. Moreover, in Gonshor approach one finds the different numbers
through the formula [40],

\begin{equation}
\mathcal{J}=n\mid \varepsilon \mid +\frac{\mid \varepsilon _{0}\mid }{2}%
+\sum \limits_{i=1}^{q}\frac{\mid \varepsilon _{i}\mid }{2^{i+1}},
\label{32}
\end{equation}%
where $\varepsilon ,\varepsilon _{0},\varepsilon _{1},...,\varepsilon
_{q}\in \{+,-\}$ and $\varepsilon \neq \varepsilon _{0}$. Furthermore, one
has $\mid +\mid =+$ and $\mid -\mid =-$. For instance, one has 
\begin{equation}
(++-+-+)=2-\frac{1}{2}+\frac{1}{4}-\frac{1}{8}+\frac{1}{16}=\frac{27}{16}.
\label{33}
\end{equation}%
A connection with $N$-qubit structure can be obtained by introducing a
surreal complex numbers $\mathcal{Z}$ in the form

\begin{equation}
\mathcal{Z}=\mathcal{J}_{1}+i\mathcal{J}_{2},  \label{34}
\end{equation}%
where $\mathcal{J}_{1}$ and $\mathcal{J}_{2}$ are two surreal numbers
according to the formula (32). So one may identify a complex operator $%
\mathcal{\hat{Z}}_{\hat{A}_{1}\hat{A}_{2}...\hat{A}_{N}}$ such that [36]%
\begin{equation}
\begin{array}{c}
\mathcal{\hat{Z}}_{\hat{A}_{1}\hat{A}_{2}...\hat{A}_{N}}\mid \hat{A}_{1}\hat{%
A}_{2}...\hat{A}_{N}>=\dsum \limits_{\hat{A}_{1},\hat{A}_{2},...,\hat{A}%
_{N}=0}^{1}Q_{\hat{A}_{1}\hat{A}_{2}...\hat{A}_{N}}\mid \hat{A}_{1}\hat{A}%
_{2}...\hat{A}_{N}> \\ 
\\ 
=\mathcal{J}\mid \hat{A}_{1}\hat{A}_{2}...\hat{A}_{N}>.%
\end{array}
\label{35}
\end{equation}%
In ordinary quantum mechanics, the $z$-component $\hat{J}_{z}$ of the total
angular momentum $\hat{J}$ is quantized as $J_{z}=n\pm \frac{1}{2}$, with
the identification of $\frac{1}{2}$-spin of the system. This means that
there must exist infinite number of $\mathcal{J}$-spins, according to the
formula (32). Thus, in general one must expect a particles with dyadic
rational $\frac{p}{2^{q}}$-spin and in particular particles with $\frac{1}{4}
$-spin (see Refs. [41] and [42]) and $\frac{1}{8}$-spin. At this respect it
is tempted to propose this $\frac{p}{2^{q}}$-spin particles as a new
candidate for dark matter.

Now, according to (11) the generalized Elko equation can be written as

\begin{equation}
\left[ \gamma _{ab}^{\mu }\hat{p}_{\mu }\psi
_{A_{1}...A_{l}}^{b}+im_{0}\delta _{ab}~^{\ast }\psi _{A_{1}...A_{l}}^{b}%
\right] =0.  \label{36}
\end{equation}%
Hence, since both qubits and surreal numbers make sense in $2^{N}$%
-dimensions one requires that a connection with the generalized Elko
equation can be established if $s=n+1$ and $2^{s}=2l$, where $s=N-n$. This
means that parameters $l$ and $N$ are linked by $2^{N-n-1}=l$ and one also
obtains $N=2n+1$. One of the simplest examples is a $3$-qubit system, with $%
Q_{\hat{A}_{1}\hat{A}_{2}\hat{A}_{3}}$ which implies the Pl\"{u}cker
coordinates $\mathcal{Q}_{A_{1}A_{2}}=\varepsilon _{i_{1}i_{2}}\mathbf{%
\omega }_{A_{1}}^{i_{1}}\mathbf{\omega }_{A_{2}}^{i_{2}}$. In turn, this
means that one must have the correspondence $\mathcal{Q}_{A_{1}A_{2}}%
\rightarrow \psi _{A_{1}A_{2}}^{a}$ and $\mathbf{\omega }_{A}^{i}\rightarrow
\psi _{A}^{ai}$, with indices in the field $\psi _{A_{1}A_{2}}^{b}$, $%
A_{1},A_{2}$...etc running from $1$ to $4$. Therefore, in this case (35)
becomes

\begin{equation}
\left[ \gamma _{ab}^{\mu }\hat{p}_{\mu }\psi _{A_{1}A_{2}}^{b}+im_{0}\delta
_{ab}~^{\ast }\psi _{A_{1}A_{2}}^{b}\right] =0,  \label{37}
\end{equation}%
with $^{\ast }\psi _{A_{1}A_{2}}^{a}=\frac{1}{2}\varepsilon
_{A_{1}A_{2}}^{~~~~~~A_{3}A_{4}}\psi _{A_{3}A_{4}}^{a}$.

From the generalized Elko equation (36) it is evident that the presence of
the quantity $^{\ast }\psi _{A_{1}...A_{l}}^{a}$ (see Eq. (11)) means that
the duality concept in terms of the $\varepsilon $-symbol ($\varepsilon $%
-duality) must be present in such an equation. In addition, we are requiring
that $\psi _{A_{1}...A_{l}}^{a}$ satisfies the complex Grassmann-Pl\"{u}cker
relations (24). Therefore, one can say that both the $\varepsilon $-duality
and the Grassmann-Pl\"{u}cker relations are part of the mathematical
structure of the generalized Elko equation. It turns out that,
mathematically, the $\varepsilon $-duality can be identified with the
Hodge-dual of $p$-forms, while Grassmann-Pl\"{u}cker relations can be
associated with the Grassmannian variety $Gr(L,l)$ corresponding to $l$%
-planes in $C^{L}$ \textit{via} the Pl\"{u}cker embedding. Therefore, both,
mathematical concepts, the Hodge-dual and the Grassmannian variety enter in
a natural way on the generalized Elko equation (36). Remarkably, these two
concepts can be considered as two basic concepts in one possible way to
define realizable oriented matroids (see Ref. [27] for details). Moreover,
if one calls $L$ the dimension of the space where the indices $%
A_{1},...,A_{l}$ `live' one observes from (11) that the dual field $^{\ast
}\psi _{A_{1}...A_{l}}^{a}$ requires the relation $L=2l$ which in turn
motivates to introduce the $L\times l$ matrix $\psi _{A}^{i}$ from which one
can obtain $\psi _{A_{1}...A_{l}}^{a}$ according to (25). Surprisingly, the
matrix $\mathbf{\psi }_{A}^{i}\equiv \psi _{A}^{ai}$ can also be obtained if
one makes the identification of the first $N-n$ and the last $n$ indices of $%
N$-qubit system $Q_{\hat{A}_{1}\hat{A}_{2}...\hat{A}_{N}}$, with the
dimensions $L$ and $l$ written in the form $L=2^{N-n}$ and $l=2^{n}$,
implying the numerical formula $N=2n+1$. Again, this is linked to the fact
that one can associate $N$-qubits with the complex Grassmannian variety $%
Gr(L,l)$ of $l$-planes in $C^{L}$ (see Ref. [37] for details). The surreal
numbers enter in this scenario via the Gonshor mathematical formalism [40]
in the sense that one can repeat exactly the same exercise that in the case
of $N$-qubit, namely identifying the the first $j-n$ and last $n$ indices of
($j$-day)-surreal number $\mathcal{J}$ with the matrix $\mathbf{\psi }%
_{A}^{i}$. However, one note that an important difference emerges when the
matrix $\mathbf{\psi }_{A}^{i}$ is associated with the ($j$-day)-surreal
number $\mathcal{J}$. This difference refers to the fact that in the case of
surreal numbers one can associate several dyadic rational type $\frac{p}{%
2^{q}}$ numbers to the matrix $\mathbf{\psi }_{A}^{i}$, where $p$ and $q$
are integer numbers. In this case, interesting possibility arises in the
sense that just as usual spinors $\psi $ can be associated with $\frac{1}{2}$%
-spin one is tempted to propose that field $\mathbf{\psi }_{A}^{i}$
corresponds to particles with $\frac{m}{2^{p}}$-spin. In particular,
surrreal numbers of the type $(++,...,+,+)$ correspond to integer spin
(bosons), while surreal numbers of the type $(++,...,+,-)$ refers to
half-integer spin (fermions). But one may have many other combinations of
plus and minus signs associated to surreal numbers. For instance, a surreal
number of the type $(+--)$ leaves to particles of $\frac{1}{4}$-spin, and so
on. Thus, one wonders whether this motivating idea can induce an alternative
route for the investigation of the origin of dark matter which, in spite of
many attempts, it is known continues as an unsolved problem.

Finally, let us just mention that in Ref. [32] (see also Ref. [29]) a
connection between fiber bundle oriented matroid theory [43] and super $p$%
-branes was established. Thus, according to the present development, one may
expect that eventually a link between super $p$-branes and the generalized
Elko theory can be considered and therefore one is tempting to believe that
our proposed generalized Elko theory may be useful in the search for quantum
gravity.

\bigskip

\begin{center}
\textbf{Acknowledgments}

{\small \ }
\end{center}

I would like to thank professor D. V. Ahluwalia, for his friendship and
making me become interested in Elko theory. I would also like to thank the
reviewer for helpful comments. This work was partially supported by
PROFAPI-UAS/2013.

\end{document}